\documentclass[english,pra,reprint]{revtex4-2}
\usepackage[T1]{fontenc}
\usepackage[utf8]{inputenc}
\usepackage{color}
\usepackage{babel}
\usepackage{mathrsfs}
\usepackage{amsmath}
\usepackage{amssymb}
\usepackage{graphicx}
\usepackage{xargs}[2008/03/08]
\usepackage[pdfusetitle,
 bookmarks=true,bookmarksnumbered=false,bookmarksopen=false,
 breaklinks=false,pdfborder={0 0 1},backref=false,colorlinks=true]
 {hyperref}
\hypersetup{
 pdfborderstyle=}

\makeatletter
\usepackage{babel}
\usepackage[justification=raggedright]{caption}

\allowdisplaybreaks[4]
\usepackage{orcidlink}

\usepackage{ragged2e}
\DeclareCaptionJustification{justified}{\justifying}
\ifdefined\showcaptionsetup
 
\fi

\ifdefined\showcaptionsetup
 \PassOptionsToPackage{caption=false}{subfig}
\fi
\usepackage{subfig}
\makeatother

\begin{document}
\title{Optimal quantum superresolution for full distance between  incoherent
optical sources  in two dimensions}
\author{Junyan Li$^{1}$\orcidlink{0009-0008-1357-6590}}
\author{Shengshi Pang$^{1,2}$\orcidlink{0000-0002-6351-539X}}
\email{pangshsh@mail.sysu.edu.cn}

\affiliation{$^{1}$School of Physics, Sun Yat-sen University, Guangzhou, Guangdong
510275, China\\$^{2}$Hefei National Laboratory, University of Science
and Technology of China, Hefei 230088, China}
\begin{abstract}
The Rayleigh criterion has long served as a fundamental limit for
the resolution of  optical imaging. Recent advances in multiparameter
quantum metrology have led to  quantum superresolution that can break
this limit and achieve nonvanishing precision in estimating the separation
between a pair of closely located incoherent point sources. For two-dimensional
optical systems, the quantum superresolution has been studied for
the Cartesian components of separation between two incoherent  point
sources. However, the precision limit of estimating the full distance
between two point sources remains unknown so far. In this paper, we
investigate the estimation precision of the full distance between
two incoherent point sources with arbitrary intensities in a two-dimensional
imaging system. Through the multiparameter quantum estimation theory,
we obtain the ultimate estimation precision for the distance and show
that it remains nonzero when the distance approaches zero, which surpasses
the Rayleigh criterion. We further show the dependence of the estimation
precision  on the relative orientation between the two point sources,
which leads to a novel scheme that can enhance the precision by aligning
the sources along proper directions if the point-spread functions
are not circularly symmetric, and the enhancement is determined by
the extent to which the point-spread functions deviate from circular
symmetry. Finally, the results are illustrated by incoherent sources
with Gaussian point-spread functions.
\end{abstract}
\maketitle
\global\long\def\hro{\hat{\rho}}%
\global\long\def\bra#1{\langle#1|}%
\global\long\def\ket#1{|#1\rangle}%
\newcommandx\ksi[1][usedefault, addprefix=\global, 1=]{\ket{\psi_{#1}}}%
\newcommandx\bsi[1][usedefault, addprefix=\global, 1=]{\bra{\psi_{#1}}}%
\newcommandx\xk[1][usedefault, addprefix=\global, 1=k]{X_{#1}}%
\newcommandx\yk[1][usedefault, addprefix=\global, 1=k]{Y_{#1}}%
\newcommandx\zk[1][usedefault, addprefix=\global, 1=k]{Z_{#1}}%
\global\long\def\dx{d_{x}}%
\global\long\def\dy{d_{y}}%
\global\long\def\dz{d_{z}}%
\global\long\def\xb{\bar{X}}%
\global\long\def\yb{\bar{Y}}%
\global\long\def\zb{\bar{Z}}%
\newcommandx\hp[2][usedefault, addprefix=\global, 1=, 2=]{\hat{P}_{#1}^{#2}}%
\newcommandx\avg[4][usedefault, addprefix=\global, 1=, 2=, 3=, 4=]{\langle\hp[#1][#4]\rangle_{#2}^{#3}}%
\global\long\def\qo{\langle\psi_{1}|\psi_{2}\rangle}%
\newcommandx\av[1][usedefault, addprefix=\global, 1=]{\langle#1\rangle}%
\global\long\def\norm#1{\left|#1\right|^{2}}%
\newcommandx\nei[2][usedefault, addprefix=\global, 1=]{e^{-i#2\hp[#1]}}%
\newcommandx\nk[1][usedefault, addprefix=\global, 1=k]{N_{#1}}%
\newcommandx\px[1][usedefault, addprefix=\global, 1=x]{\partial_{#1}}%
\newcommandx\py[1][usedefault, addprefix=\global, 1=y]{\partial_{#1}}%
\newcommandx\pz[1][usedefault, addprefix=\global, 1=z]{\partial_{#1}}%
\newcommandx\nt[1][usedefault, addprefix=\global, 1=]{N_{{\rm tot}}^{#1}}%
\newcommandx\nd[1][usedefault, addprefix=\global, 1=]{N_{{\rm diff}}^{#1}}%
\global\long\def\hl#1{\hat{\mathcal{L}}_{#1}}%
\newcommandx\gi[1][usedefault, addprefix=\global, 1=i]{g_{#1}}%
\global\long\def\var{{\rm Var}}%
\newcommandx\hr[1][usedefault, addprefix=\global, 1=]{\mathcal{H}_{r}^{#1}}%
\newcommandx\ha[1][usedefault, addprefix=\global, 1=]{\mathcal{H}_{\alpha}^{#1}}%
\global\long\def\hrn{\mathcal{H}_{r}^{({\rm num})}}%
\global\long\def\hrd{\mathcal{H}_{r}^{({\rm den})}}%
\newcommandx\ei[2][usedefault, addprefix=\global, 1=]{e^{i#2\hp[#1]}}%
\global\long\def\dlm#1{\left.#1\right|_{r\rightarrow0}}%
\global\long\def\hrop{\mathcal{H}_{r}^{({\rm opt})}}%
\global\long\def\hroc{\mathcal{H}_{r}^{({\rm cir})}}%
\global\long\def\ap{\alpha_{{\rm opt}}}%
\global\long\def\aw{\alpha_{{\rm wor}}}%

\section{Introduction}

In recent years, quantum metrology has been rapidly advanced across
a range of fields, including gravitational wave detection \citep{Adhikari2014,Roura2020},
quantum clocks \citep{Ludlow2015,Roura2020,Woods2022,Song2013}, quantum
sensing \citep{Degen2017a,Pirandola2018}, quantum imaging \citep{Lemos2014,Moreau2019,Perez-Delgado2012}
and provided new perspectives for the quantum-enhanced technologies
that can boost the performance of measurements beyond the precision
limit of their classical counterparts.

A notable example is the longstanding challenge of resolving closely
spaced optical sources, which is constrained by Rayleigh criterion
in classical optics \citep{rayleighXXXIInvestigationsOptics1879}.
This criterion sets a diffraction-limited resolution threshold at
$d=1.22\lambda/(2{\rm NA})$, with $\lambda$ the wavelength of the
light and ${\rm NA}$ the numerical aperture of the imaging system.
To overcome this limit, various approaches have been proposed. One
straightforward method is to reduce the wavelength or increase the
numerical aperture \citep{Schropp2012}. More fundamentally, techniques
such as microscopy structured illumination \citep{Gustafsson2005,Rittweger2009},
photoactivatable localization \citep{Betzig2006} and stochastic optical
reconstruction microscopy \citep{Rust2006} have been proposed to
surpass the Rayleigh limit by exploiting nonlinear optical responses,
stochastic activation of fluorescent emitters, and computational reconstruction
algorithms. While these techniques have achieved remarkable progress
in imaging resolution, they typically require specialized fluorophores,
sophisticated control of activation cycles, or complex image processing
\citep{Weisenburger2015a}.

From a statistical perspective, when two incoherent optical sources
have significant overlap on the image plane, the precision of estimating
their separation is fundamentally limited by the Cramér-Rao lower
bound \citep{Kay1993,Zmuidzinas2003}. It follows that the estimation
precision of the separation between two optical sources vanishes as
the separation approaches zero by direct imaging. This statistical
phenomenon is known as Rayleigh’s curse. However, Tsang et al. \citep{Tsang2016}
attacked this problem from the perspective of quantum multiparameter
estimation and surprisingly found that the quantum Cramér-Rao bound
(QCRB) \citep{Helstrom1969} does not vanish for arbitrarily small
separation of two incoherent point sources if the quantum measurements
on the point sources are optimized, which addresses Rayleigh's curse.

Since Tsang et al.'s breakthrough, the quantum superresolution has
been verified experimentally by various techniques, e.g., image inversion
interferometry \citep{Tang2016b}, heterodyne detection \citep{Yang2016},
edge coherence inversion \citep{Tham2017}, two-photon interference
\citep{Parniak2018,Thachil2023}, and digital holography \citep{Paur2016}.
In parallel, theoretical advances have obtained the optimal observables
and estimators for the quantum superresolution \citep{Sorelli2021,Sorelli2021a}
and extended the quantum superresolution framework to thermal states
\citep{Nair2016}, detector noise \citep{Lupo2020a,Tsang2023}, dark
states \citep{Brennan2022}, arbitrary source distributions \citep{Tsang2021a,Matlin2022},
moment estimation \citep{Tan2023,Tsang2019,Zhou2019}. Further developments
have considered the influence of unknown intensities of the optical
sources which leads to the resurgence of Rayleigh's curse \citep{Rehacek2017},
and studied the roles of coherence \citep{Hradil2019,Liang2021a,Zheltikov2022,Karuseichyk2022,Liang2021}
and partial coherence \citep{Larson2018,Tsang2019c,Larson2019a,Wadood2021,Liang2023}
of optical sources in preventing the resurgence of Rayleigh's curse.
The feasibility conditions of superresolution given unknown intensities
of the point sources have also been obtained \citep{Li2024}. These
efforts have also enabled the application of quantum superresolution
in source localization \citep{Prasad2020a,Bisketzi2019}, time-frequency
resolution \citep{Niewelt2023,Ansari2021,Mazelanik2022,mitchellQuantumLimitsResolution2022,Donohue2018},
etc.

In the original superresolution protocol \citep{Tsang2016}, the superresolution
was proposed for two optical sources with one-dimensional point-spread
functions . However, the point-spread functions of practical optical
sources are essentially multidimensional, where the spatial structures
of the point-spread functions become crucial to the imaging resolution.
In this broader context, multiparameter quantum estimation theory
provides a powerful tool to improve the imaging precision \citep{Ragy2016,Chrostowski2017,Yang2019,Hervas2024}.
Building upon this foundation, subsequent researches have extended
the quantum superresolution to the estimation of Cartesian components
of the separation for two-dimensional point sources \citep{Ang2017}
and for three-dimensional point sources \citep{Yu2018,Prasad2020,Wang2021a},
and to more complex imaging tasks such as surface metrology \citep{Napoli2019},
using balanced homodyne detection \citep{Gosalia2024}, and estimating
the sizes of uniform line and disk sources \citep{Prasad2020b}, etc.

In this work, we investigate the quantum precision limits of two-dimensional
superresolution for a pair of incoherent point sources with arbitrary
relative intensities. Rather than the individual Cartesian components
of the separation, we focus on the quantum estimation of the full
distance between two point sources. We show that, by optimizing the
measurement basis, the quantum Fisher information for the distance
remains finite when the two sources get close to each other, which
breaks the Rayleigh limit. Furthermore, when the point-spread functions
(PSF) of the two point sources are not circularly symmetric, the estimation
precision exhibits a nontrivial dependence on the \textit{relative}
orientation between the two point sources. This orientation dependence
enables further precision enhancement by aligning the two point sources
along specific directions. We optimize the quantum Fisher information
of the distance over the azimuth between the two point sources, and
obtain the optimal azimuth and the maximum quantum Fisher information
for the estimation of source distance. Taking Gaussian PSFs as an
example, we find that the optimal orientation corresponds to the minor
axis of the PSF.

\section{Preliminaries}

\subsection{Model of optical sources}

For a spatially invariant imaging system, suppose two incoherent optical
pointlike sources are located at unknown positions $(\xk[1],\yk[1])$
and $(\xk[2],\yk[2])$ on the object plane, and separated by a distance
$r$. In the Poisson limit, the probability of multiple photons arriving
at the imaging plane simultaneously is negligible \citep{Chrostowski2017},
within a given coherence time interval. The average state of a single
photon that arrived at the imaging plane can be described by a density
operator \citep{Mandel1959}, 
\begin{equation}
\hro=\frac{1-\epsilon}{2}\ksi[1]\bsi[1]+\frac{1+\epsilon}{2}\ksi[2]\bsi[2],\label{eq:rho}
\end{equation}
where $\epsilon$ is the intensity imbalance parameter 
\begin{equation}
\epsilon=\frac{N_{2}-N_{1}}{N_{2}+N_{1}},\label{eq:eps}
\end{equation}
and $N_{1}$ and $N_{2}$ are the average photon numbers emitted by
the two point sources respectively. $\ksi[1,2]$ are the spatially
shifted states from a fixed state $\ksi$, given by 
\begin{equation}
\ksi[1,2]={\rm exp}(-iY_{1,2}\hp[y]-iX_{1,2}\hp[x])\ksi,\label{eq:psi}
\end{equation}
where $\hp[k]=-i\partial_{k}$, $k=x,y$, are the momentum operators
in the $x,y$ directions respectively, and the state $\ksi$ is 
\begin{equation}
\ksi=\int_{-\infty}^{\infty}dy\int_{-\infty}^{\infty}dx\psi(x,y)\ket{x,y},
\end{equation}
where $\psi(x,y)$ is the point-spread function of the imaging system,
assumed to be real and inversion-symmetric throughout this work, i.e.,
$\psi(x,y)=\psi(-x,-y)$.

\begin{figure}[!h]
\centering{}\includegraphics[scale=0.25]{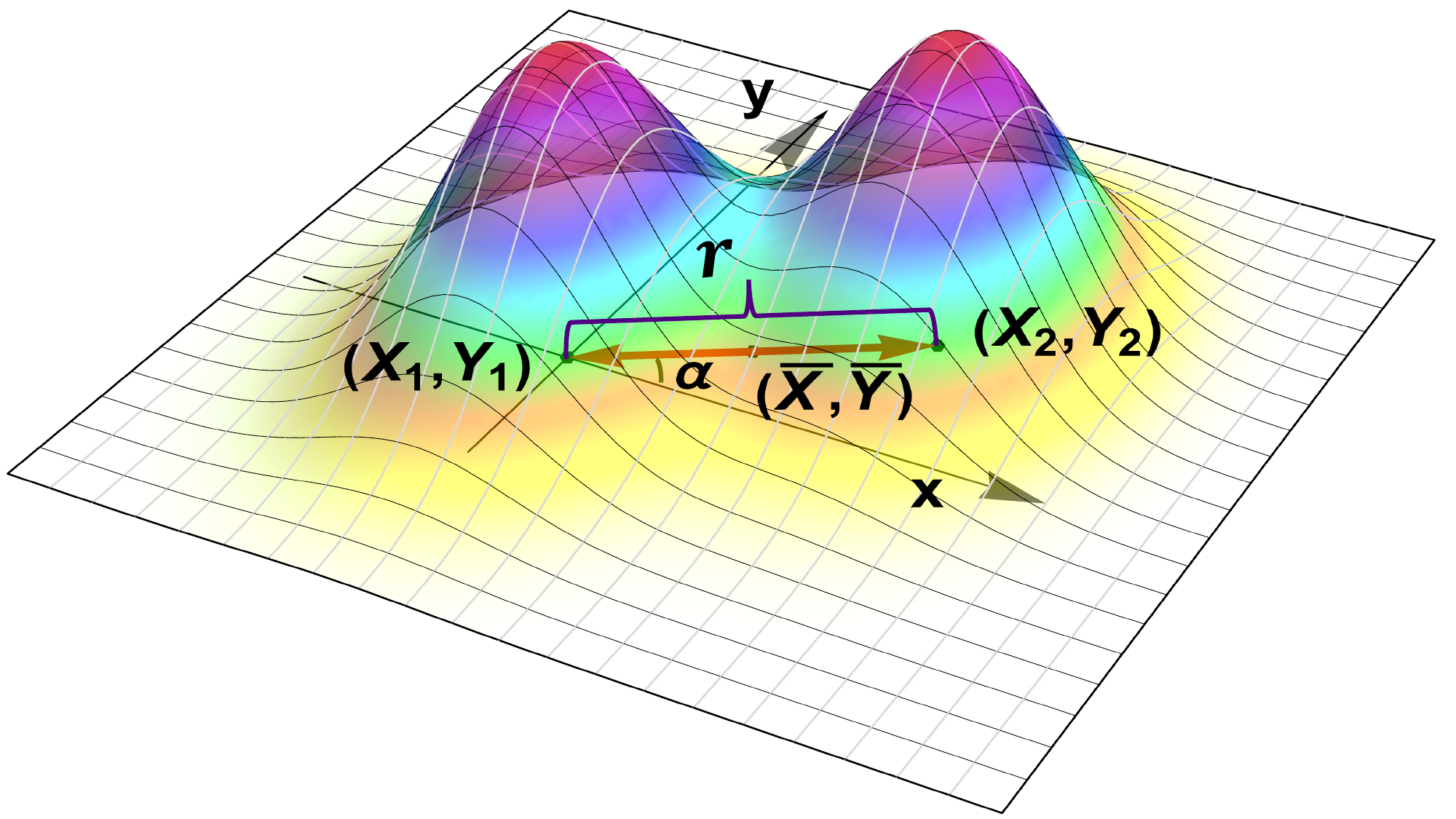}\caption{Two point optical sources located at $(X_{1},Y_{1})$ and $(X_{2},Y_{2})$.
The distance between the two sources is $r$, and the \emph{relative}
orientation between the two point sources characterized by the azimuth
$\alpha$, i.e., the angle between the displacement of the two sources
and a given reference axis which is the $x$ axis in the figure.}\label{fig:1}
\end{figure}

For two-dimensional imaging systems, in addition to the distance between
two point sources, the relative orientation also play a critical role
in the imaging resolution, so we will focus on both the distance and
the azimuth between two point sources in this work. To this end, we
define the distance between two point sources as $r$ and the azimuth
between the sources, i.e., the angle between the displacement of the
sources and a fixed reference axis, as $\alpha\in[0,\pi]$, which
is illustrated in Fig. \ref{fig:1}. The positions of the two sources
can then be written as 
\begin{equation}
(\xk[1,2],\yk[1,2])=\left(\xb\mp\frac{r}{2}{\rm cos}\alpha,\yb\mp\frac{r}{2}{\rm sin}\alpha\right),\label{eq:xy}
\end{equation}
where the subscript $1,2$ and the $\mp$ sign denote the two sources
respectively, and the centroid vector $(\bar{X},\bar{Y})$ is defined
as 
\begin{equation}
(\bar{X},\bar{Y})=(\frac{X_{2}+X_{1}}{2},\frac{Y_{2}+Y_{1}}{2}).
\end{equation}
The displacement between the two point sources can be denoted as 
\begin{equation}
(d_{x},d_{y})=(X_{2}-X_{1},Y_{2}-Y_{1}).
\end{equation}
In this case, the momentum operator along the displacement between
the two sources can be defined as 
\begin{equation}
\hp[r]=\hp[x]\cos\alpha+\hp[y]\sin\alpha,
\end{equation}
with its orthogonal complement given by 
\begin{equation}
\hp[r^{\perp}]=-\hp[x]\sin\alpha+\hp[y]\cos\alpha.
\end{equation}

\begin{figure}[!h]
\subfloat[\centering$\alpha=0$]{\centering{}\includegraphics[scale=0.33]{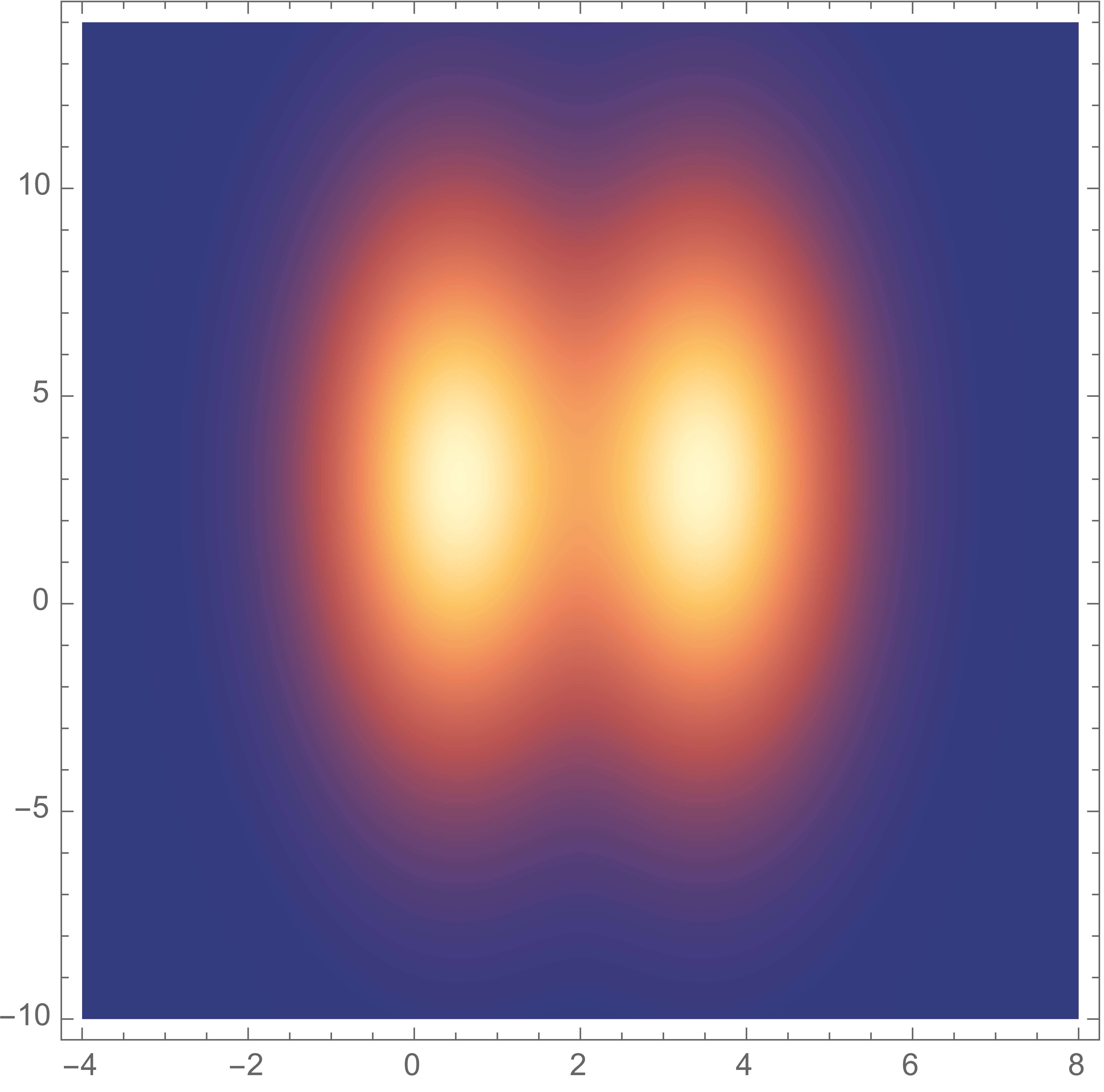}

}\medspace{}\subfloat[\centering$\alpha=\frac{\pi}{2}$]{\begin{centering}
\includegraphics[scale=0.33]{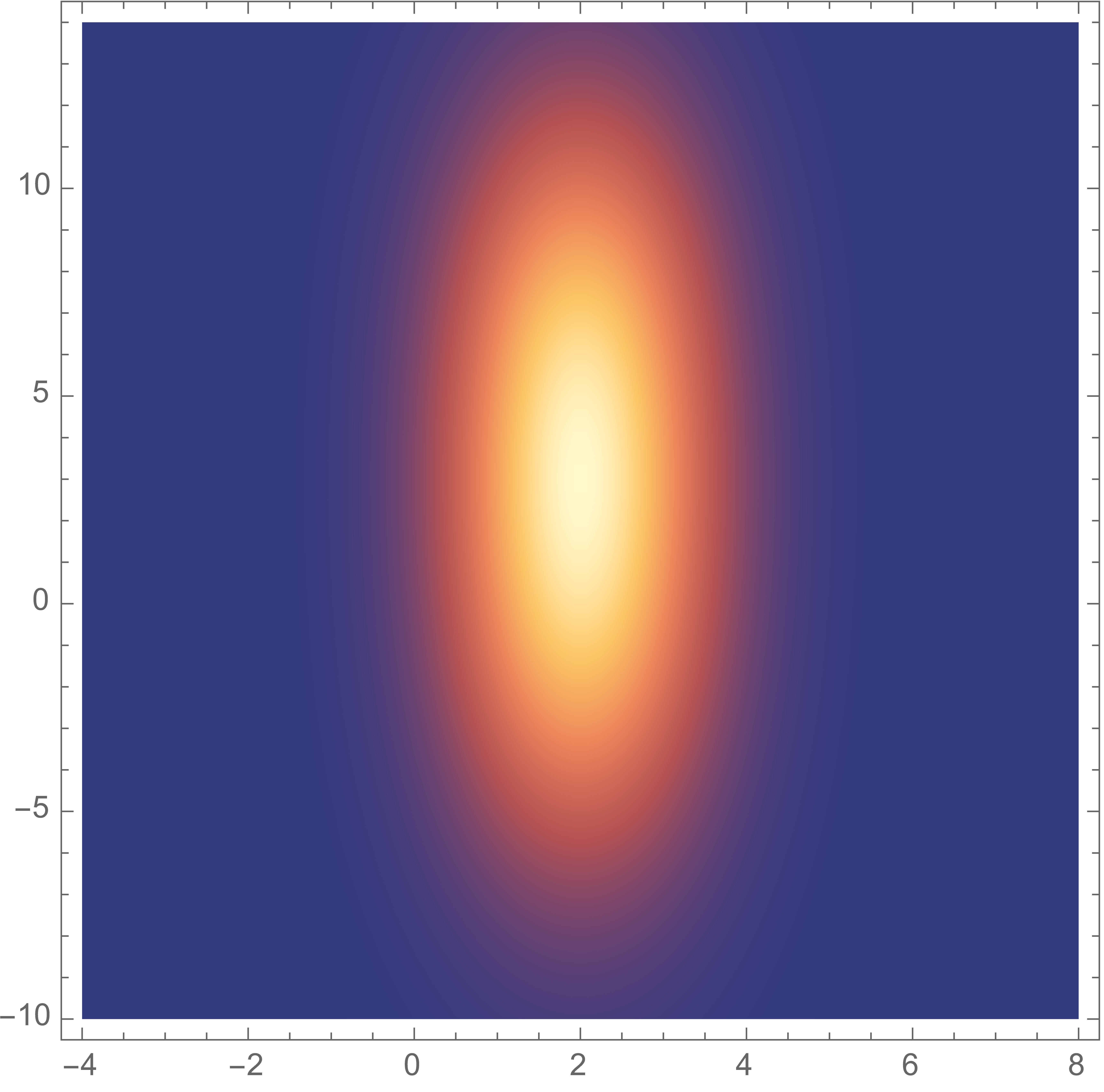} 
\par\end{centering}
\centering{}

}\caption{The probability distribution $p(x,y)$ for detecting a single photon
from two point sources with identical Gaussian point-spread functions
on a two-dimensional image plane with respect to the different azimuth
$\alpha$. The plots show how the spatial overlap $\delta$ between
the two point-spread functions and the probability distribution of
photon detection varies with the azimuth $\alpha$.}\label{fig:a}
\end{figure}

It is worth noting that while the azimuth $\alpha$ is defined with
respect to a specific reference axis, it characterizes the \emph{relative}
orientation between the two point sources which is intrinsic to the
two point sources. This is manifested by the fact that the overlap
between the two point sources, 
\begin{equation}
\delta=\langle\psi_{1}|\psi_{2}\rangle=\av[{\cos(r\hp[r])}],
\end{equation}
is dependent on the azimuth $\alpha$ with respect to a given axis,
as illustrated in Fig. \ref{fig:a}. So, even though the azimuth $\alpha$
can change with the reference axis, the relative orientation it defines
as well as the impact of the relative orientation on the resolution
of the two point sources does not change. This will be clearer when
we obtain the quantum Fisher information for the distance later.

As shown in Fig. \ref{fig:a}, the overlap between the point-spread
functions of the two optical sources leads to the non-orthogonality
of their associated quantum states, which in turn constrains the distinguishability
between them. In particular, when the distance is comparable to or
even smaller than the width of the point-spread function, the two
point sources becomes essentially indistinguishable, which is known
as Rayleigh's curse. This raises a challenging question: how can one
estimate the distance between two incoherent point sources, particularly
when the distance goes to zero?

This question has been addressed for one-dimensional imaging systems
by the pioneering work of Tsang et al., and also been considered for
the estimation of the orthogonal components of the separation between
two point sources in two-dimensional imaging systems, as reviewed
in the introduction section. In this work, we will focus on the estimation
of the full distance, as well as the relative orientation, between
two point sources in two-dimensional imaging systems.

\subsection{Precision limit of direct imaging}

The most straightforward measurement scheme to resolve two closely
located point sources is the direct imaging, wherein the information
about the positions of the point sources can be extracted from the
intensity distribution of the photons arriving at the image plane.

Suppose the photon counting at a small area $dx\times dy$ around
the position $(x,y)$ of the image plane is approximately $(N_{1}+N_{2})p_{{\bf g}}(x,y){\rm dx}{\rm dy}$,
where $p_{{\bf g}}(x,y)$ is the position distribution of a single
photon arriving on the image plane, 
\begin{equation}
p_{{\bf g}}(x,y)=\frac{1-\epsilon}{2}\Lambda(x-X_{1},y-Y_{1})+\frac{1+\epsilon}{2}\Lambda(x-X_{2},y-Y_{2}),\label{eq:dp}
\end{equation}
and it depends on four parameters collected in the vector ${\bf g}$:
the distance $r$, the azimuth $\alpha$ and the centroid positions
$\left(\xb,\yb\right)$, i.e., 
\begin{equation}
{\bf g}=(\xb,\yb,r,\alpha)^{\top}.\label{eq:para}
\end{equation}
The intensity $\Lambda(x,y)$ is determined by the point-spread function
$\psi(x,y)$ of the imaging system, 
\begin{equation}
\Lambda(x,y)=|\psi(x,y)|^{2}.
\end{equation}
To resolve two optical point sources, it is essential to measure and
estimate the parameter vector ${\bf g}$.

For any unbiased estimator ${\bf \hat{g}}=(\widehat{\xb},\widehat{\yb},\widehat{r},\widehat{\alpha})^{\intercal}$,
the covariance matrix of the estimation is lower bounded by the well-known
Cramér-Rao inequality \citep{Helstrom1969,Yang2019a,Liu2020}, 
\begin{equation}
{\rm Cov}[{\bf \hat{g}}]\geq\nt[-1]\mathcal{J}^{-1}[{\bf g}],\label{eq:cramer-rao}
\end{equation}
where $\nt$ is total number of photons from the two optical sources,
\begin{equation}
\nt=N_{1}+N_{2}.\label{eq:ntot}
\end{equation}
The ``$\geq$'' sign represents semi-definite positivity of matrix,
and $\mathcal{J}[{\bf g}]$ is the Fisher information matrix per detected
photon. When the photon numbers $\nt$ are sufficiently large, the
Cramér-Rao bound inequality can always be asymptotically achieved
\citep{Helstrom1969}.

For a small distance $r$, the estimation precision using the direct
imaging method is given by 
\begin{equation}
\mathcal{H}_{r}^{({\rm direct})}=\nt/(\mathcal{J}^{-1})_{33}=\frac{r^{2}\left(1-\epsilon^{2}\right)^{2}\nt}{32}\mathcal{A},
\end{equation}
as derived in Appendix \ref{sec:Direct-image}, where $\mathcal{A}$
is independent of the distance $r$. This result implies that, by
direct imaging, the estimation precision $\mathcal{H}_{r}^{({\rm direct})}$
decays quadratically with $r$ and thus vanishes in the limit $r\rightarrow0$.
This phenomenon is called Rayleigh's curse \citep{Tsang2016}.

\section{Superresolution for two-dimensional point sources}

\subsection{Quantum precision limit for separation of point sources}

Although direct imaging is a standard imaging method which is straightforward
to implement, it is only one of the possible measurement methods allowed
by quantum mechanics.

For the current quantum superresolution problem, the quantum state
of a single photon is described by the density operator $\hro$ \eqref{eq:rho},
which depends on a set of unknown parameters ${\bf g}=(\xb,\yb,r,\alpha)^{\top}$
\eqref{eq:para}. A brief review of quantum estimation theory in the
context of this problem is provided in Appendix \ref{subsec:Derivation=00003D000020f},
where it is shown that the estimation precision of the parameter vector
${\bf g}$ is fundamentally constrained by the quantum Cramér--Rao
bound (QCRB), 
\begin{equation}
{\rm Cov}[{\bf \hat{g}}]\geq\nt[-1]\mathcal{Q}_{ij}^{-1}[{\bf g}],
\end{equation}
where $\mathcal{Q}$ is the quantum Fisher information matrix, determined
by the quantum state $\hro$ to be measured.

In fact, the quantum Cramér-Rao bound for multiparameter estimation
is not always achievable due to the potential non-commutativity between
the optimal measurements for different parameters. But one can introduce
a positive-semidefinite weight matrix $W$ to assign different weights
with different parameters, yielding a scalar bound 
\begin{equation}
{\rm Tr}(WC)\geq\nt[-1]{\rm Tr}(W\mathcal{Q}^{-1}).\label{eq:scalar}
\end{equation}
This scalar bound is saturable when 
\begin{equation}
{\rm Tr}(\hro[\hl{g_{i}},\hl{g_{j}}])={\rm ImTr}(\hro\hl{g_{i}}\hl{g_{j}})=0,\label{eq:compatibility}
\end{equation}
where $\hl{g_{i}}$ denotes the symmetric logarithmic derivative (SLD)
with respect to parameter $g_{i}$, since the quantum Cramér-Rao bound
coincides with the Holevo bound when this condition holds and the
latter is always asymptotically achievable with a sufficiently large
number of states \citep{Ragy2016,Rehacek2017}. In our model, this
condition is fulfilled due to the reality of $\psi(x)$, which makes
the symmetric logarithmic derivatives $\hl{g_{i}}$ and $\hl{g_{j}}$
also real and hence ${\rm Tr}(\hro\hl{g_{i}}\hl{g_{j}})$ real as
well. Therefore, the quantum Cramér-Rao bound provides an achievable
precision limit to quantum superresolution.

By diagonalizing $\hro$, the quantum Fisher information matrix with
respect to the four unknown parameters $\xb,\yb,r,\alpha$ can be
obtained straightforwardly by its definition. As the derivation and
the result are lengthy, they are provided in Appendix \ref{subsec:2dqf}.

In the limit $r\rightarrow0$, the precision $\hr$ is simplified
to 
\begin{equation}
\dlm{\hr}=\frac{\text{\ensuremath{\nt}}\left(1-\epsilon^{2}\right)\left(\kappa_{x}\kappa_{y}-\eta^{2}\right)}{\kappa_{r^{\perp}}},\label{eq:hds}
\end{equation}
where 
\begin{equation}
\eta=\av[{\hp[x]\hp[y]}],\;\kappa_{i}=\avg[i][][][2],i=x,y,r^{\perp},
\end{equation}
and $r^{\perp}$ is the direction orthogonal to $r$. This result
demonstrates that, while direct imaging fails with vanishing precision
in the sub-Rayleigh regime, an optimized quantum measurement scheme
can achieve non-vanishing precision, thus beating Rayleigh’s curse.

Notably, as aforementioned, the azimuth $\alpha$ between the two
point sources is defined with respect to a specific reference axis,
which can be changed with the coordinating system, but the relative
orientation between the two point sources is inherent to the two sources
and the impact of the relative orientation on the superresolution
precision is independent of the choice of coordinate system. This
is particularly manifested by the result $\hr$. While $\hr$ is dependent
on the azimuth $\alpha$, the precision $\hr$ turns out to be invariant
with any rotation of the coordinate system. A detailed proof for this
coordinate invariance of $\hr$ is provided in Appendix \ref{subsec:The-Coordinate-Invariance}.

It can be immediately seen from Eq. \eqref{eq:hds} that the estimation
precision $\hr$ attains its maximum over $\epsilon$ when the intensities
of two optical sources are balanced, i.e., $\epsilon=0$. In this
case, the precision $\mathcal{H}_{r}$ \eqref{eq:hds} can be further
simplified to 
\begin{equation}
\dlm{\hr}=\frac{\text{\ensuremath{\nt}}\left(\kappa_{x}\kappa_{y}-\eta^{2}\right)}{\kappa_{r^{\perp}}}.
\end{equation}
Any deviation from this balance leads to a degradation in precision.
Counterintuitively, this suggests that introducing asymmetry to the
intensities of two sources, while increasing the visual contrast of
the intensity distribution on the image plane, actually reduces the
resolvability of the two sources. Physically, this is because the
resolution of two optical sources relies on the distinguishability
between them and lowering the intensity of either source will decrease
its distinguishability from the other. For example, if the intensity
of one source is tuned to zero, the visual contrast on the image plane
reaches the maximum, but that source can never be distinguished from
the other as it is impossible to determine its location when its intensity
is zero.

\subsection{Optimization for precision limit of distance}

As explicitly seen from Eq. \eqref{eq:hds}, the estimation precision
$\mathcal{H}_{r}$ depends sensitively on the azimuth $\alpha$, leading
to an interesting question: Can the precision $\hr$ be enhanced through
properly aligning the two sources?

We derive the optimal azimuth and the highest precision by solving
\begin{equation}
\partial\dlm{\hr}/\partial\alpha=0,
\end{equation}
where $\dlm{\hr}$ is given in Eq. \eqref{eq:hds}. The optimal and
the worst azimuths can worked out as 
\begin{equation}
\alpha_{{\rm opt/wor}}=\tan^{-1}\left(\frac{\kappa_{y}-\kappa_{x}\pm\sqrt{4\eta^{2}+\left(\kappa_{y}-\kappa_{x}\right)^{2}}}{2\eta}\right),
\end{equation}
and the corresponding highest and lowest precision in the limit $r\rightarrow0$
reads 
\begin{equation}
\begin{aligned}\dlm{\hr[({\rm opt/wor})]}= & \frac{\nt\left(1-\epsilon^{2}\right)}{2}\\
 & \times\left(\kappa_{x}+\kappa_{y}\pm\sqrt{4\eta^{2}+\left(\kappa_{y}-\kappa_{x}\right)^{2}}\right).
\end{aligned}
\label{eq:hw}
\end{equation}
The precision enhancement of properly aligning the two point sources
can be characterized by the ratio 
\begin{equation}
\xi=\frac{\hr[(\ap)]}{\hr[(\aw)]}=\frac{\kappa_{x}+\kappa_{y}+\sqrt{4\eta^{2}+\left(\kappa_{y}-\kappa_{x}\right)^{2}}}{\kappa_{x}+\kappa_{y}-\sqrt{4\eta^{2}+\left(\kappa_{y}-\kappa_{x}\right)^{2}}}.
\end{equation}
Whenever the point-spread function is not circularly symmetric, $\kappa_{y}\neq\kappa_{x}$
and $\sqrt{4\eta^{2}+\left(\kappa_{y}-\kappa_{x}\right)^{2}}>0$,
hence it indicates that optimal alignment of the two point sources
can enhances the superresolution precision significantly.

In particular, when the point-spread function of the imaging system
has reflection symmetry, e.g., $\psi(x,y)=\psi(x,-y)=\psi(-x,-y)$,
the term $\eta=\av[{\hp[x]\hp[y]}]$ vanishes \citep{Ang2017}, and
the estimation precision $\hr$ \eqref{eq:hds} is simplified to 
\begin{equation}
\dlm{\hr}=\frac{\text{\ensuremath{\nt}}\kappa_{x}\kappa_{y}}{\sin^{2}\alpha\kappa_{x}+\cos^{2}\alpha\kappa_{y}},
\end{equation}
with the optimal azimuth as 
\begin{equation}
\ap=\begin{cases}
0, & {\rm if}\,\kappa_{x}>\kappa_{y},\\
\frac{\pi}{2}, & {\rm if}\,\kappa_{x}<\kappa_{y},
\end{cases}
\end{equation}
and the worst azimuth as 
\begin{equation}
\aw=\ap+\frac{\pi}{2}.
\end{equation}
In this scenario, the highest and the lowest precision reads 
\begin{equation}
\hr[({\rm opt/wor})]=\left(1-\epsilon^{2}\right)\nt\kappa_{\max/\min},
\end{equation}
where $\kappa_{\max}$ and $\kappa_{\min}$ are the maximum and the
minimum of $\kappa_{x}$ and $\kappa_{y}$ respectively, and the ratio
$\xi$ between the highest and the lowest precisions becomes 
\begin{equation}
\xi=\frac{\kappa_{\max}}{\kappa_{\min}},
\end{equation}
implying that the more the point-spread function deviates from the
circular symmetry, the higher the superresolution precision can be
increased by proper aligning of the two point sources.

\subsection{Example: two-dimensional Gaussian point-spread functions}

To illustrate above results, we consider a typical Gaussian point-spread
functions for the imaging system, 
\begin{equation}
\begin{aligned}\psi_{i}(x,y)= & \frac{\exp\left[-\frac{1}{4}\left(f-\mu_{i}\right)^{{\rm \top}}\Sigma^{-1}\left(f-\mu_{i}\right)\right]}{(2\pi)^{1/2}\left|\Sigma\right|^{1/4}},\;i=1,2,\end{aligned}
\label{eq:gaussian}
\end{equation}
where $f=(x,y)^{\top}$is the spatial coordinate vector, and $\mu_{i}=(\xk[i],\yk[i])^{\top}$denotes
the position of the $i$-th source. The covariance matrix $\Sigma$
is defined as 
\begin{equation}
\Sigma=\left(\begin{array}{cc}
\sigma_{1}^{2} & \beta\sigma_{1}\sigma_{2}\\
\beta\sigma_{2}\sigma_{1} & \sigma_{2}^{2}
\end{array}\right),\label{eq:cov}
\end{equation}
where $\sigma_{1}$ and $\sigma_{2}$ are the widths of the Gaussian
distribution along the $x$ and $y$ axis respectively, and $\beta\in[-1,1]$
quantifies the correlation.

The spatial characteristics of a Gaussian point-spread function is
fully captured by its covariance matrix $\Sigma$ \eqref{eq:cov},
whose eigenvectors determine the orientations of the major and minor
axes, 
\begin{equation}
\Phi_{{\rm \pm}}=\tan^{-1}\frac{\sigma_{2}^{2}-\sigma_{1}^{2}\pm\Re}{2\beta\sigma_{1}\sigma_{2}},\label{eq:axis}
\end{equation}
where $\Re\equiv\sqrt{(\sigma_{2}^{2}-\sigma_{1}^{2})^{2}+4\beta^{2}\sigma_{2}^{2}\sigma_{1}^{2}}$,
and $\Phi_{\pm}$ are the angles of the major and minor axes with
respect to the horizontal axis, respectively. Additionally, the quantities
in Eq. \eqref{eq:hds} become 
\begin{equation}
\begin{aligned}\kappa_{x}=\frac{1}{4\sigma_{1}^{2}\left(1-\beta^{2}\right)},\\
\kappa_{y}=\frac{1}{4\sigma_{2}^{2}\left(1-\beta^{2}\right)},\\
\eta=-\frac{\beta}{4\sigma_{1}\sigma_{2}\left(1-\beta^{2}\right)}.
\end{aligned}
\end{equation}

For two Gaussian point sources with arbitrary intensities, the estimation
precision of the distance $r$ in the limit $r\rightarrow0$ is 
\begin{equation}
\dlm{\hr}=\frac{\nt\left(1-\epsilon^{2}\right)}{4\left(\beta\sigma_{2}\sigma_{1}\sin(2\alpha)+\sigma_{1}^{2}\cos^{2}\alpha+\sigma_{2}^{2}\sin^{2}\alpha\right)},\label{eq:hdg}
\end{equation}
it can be seen that the estimation precision $\mathcal{H}_{r}$ remains
nonzero in the sub-Rayleigh limit and shows its dependence on the
azimuth $\alpha$.

The extrema of the precision $\mathcal{H}_{r}$ over the azimuth $\alpha$
can be determined by the stationary condition 
\begin{equation}
\partial\hr/\partial\alpha=0,
\end{equation}
which gives the optimal and the worst $\alpha$ as 
\begin{equation}
\alpha_{{\rm opt/wor}}=\tan^{-1}\frac{\sigma_{2}^{2}-\sigma_{1}^{2}\mp\Re}{2\beta\sigma_{1}\sigma_{2}}.
\end{equation}
One can observe that $\ap=\Phi_{{\rm -}}$ and $\aw=\Phi_{+}$, implying
that the extrema of the estimation precision $\hr$ occurs when the
displacement between the two point sources aligns with one of the
principal axes of the Gaussian point-spread function. The optimal
and the worst precision $\hr$ as $r\rightarrow0$ are given by 
\begin{equation}
\dlm{\hr[(\alpha_{{\rm opt/wor}})]}=\frac{\nt\left(1-\epsilon^{2}\right)\left(\sigma_{1}^{2}+\sigma_{2}^{2}\pm\Re\right)}{8\left(1-\beta^{2}\right)\sigma_{1}^{2}\sigma_{2}^{2}},\label{eq:hwg}
\end{equation}
which indicates that the optimal estimation precision $\hr$ is achieved
when the displacement between the two point sources aligns with the
minor axis of the Gaussian point-spread function.

The enhancement ratio $\xi$ by optimizing the aligning of the two
point sources in this case is given by 
\begin{equation}
\xi=\frac{\sigma_{1}^{2}+\sigma_{2}^{2}+\Re}{\sigma_{1}^{2}+\sigma_{2}^{2}-\Re}.\label{eq:de}
\end{equation}
\begin{figure}[!th]
\begin{centering}
\includegraphics[scale=0.6]{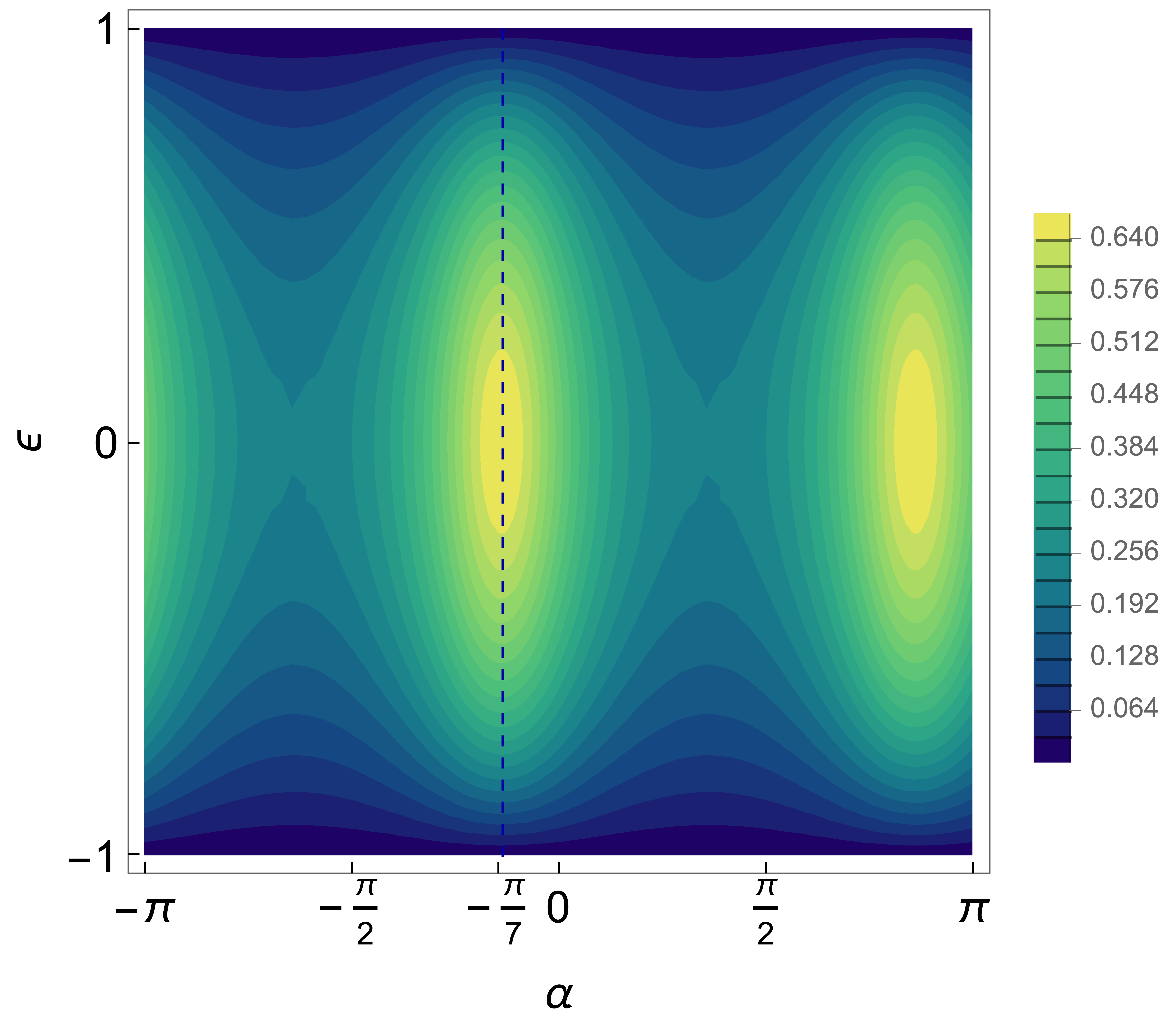} 
\par\end{centering}
\caption{Density plot of the distance precision $\protect\hr$ for two Gaussian
point-spread functions with respect to the azimuth $\alpha$ and and
the ratio $\epsilon$ of the photon number difference to the total
photon number as $r\rightarrow0$. $\sigma_{1}=1$, $\sigma_{2}=1.4$
and $\beta=0.4$. It can be observed that when the azimuth $\alpha$
is fixed, the estimation precision of the distance $r$ will decrease
as the ratio $\epsilon$ increases and the optimal estimation precision
can be always reached when the photon numbers of the two sources are
the same, i.e., $\epsilon=0$, showing the negative impact of photon
number difference on the superresolution precision. On the other hand,
when $\epsilon$ is fixed, the estimation precision of the distance
varies with the azimuth $\alpha$, and the maximum precision of the
distance $r$ can be obtained when $\alpha$ is approximately $-\pi/7$,
which corresponds to the direction along the minor axis of the Gaussian
point-spread function, implying the precision of estimating the distance
can be enhanced by optimizing the relative orientation between the
two point sources.}\label{fig:opta}
\end{figure}

The above results can be further simplified for a Gaussian point-spread
function with reflection symmetry about the $x$ and $y$ axes, i.e.,
$\beta=0$ and $\Phi_{+}=0,\;\Phi_{-}=\frac{\pi}{2}$. In this case,
the estimation precision $\hr$ \eqref{eq:hdg} becomes 
\begin{equation}
\dlm{\hr}=\frac{\left(1-\epsilon^{2}\right)\nt}{4\left(\sigma_{2}^{2}\sin^{2}\alpha+\sigma_{1}^{2}\cos^{2}\alpha\right)}.
\end{equation}
The optimal azimuth $\alpha$ for $\hr$ is 
\begin{equation}
\ap=\begin{cases}
\frac{\pi}{2}, & {\rm if}\;w<\sigma,\\
0, & {\rm if}\;\sigma<w,
\end{cases}
\end{equation}
and the highest precision turns out to be 
\begin{equation}
\max\dlm{\hr}=\frac{\left(1-\epsilon^{2}\right)\nt}{4\sigma_{\min}^{2}}.
\end{equation}
Similarly, the worst azimuth is 
\begin{equation}
\aw=\ap+\frac{\pi}{2},
\end{equation}
and the lowest precision is 
\begin{equation}
\min\dlm{\hr}=\frac{\left(1-\epsilon^{2}\right)\nt}{4\sigma_{\max}^{2}}.
\end{equation}

These results confirm that the maximum estimation precision is attained
when the two point sources are aligned along the minor axis of the
Gaussian point-spread function, in accordance with the general analysis
above. The maximum enhancement by the alignment of the point sources
is therefore 
\begin{equation}
\xi=\frac{\sigma_{\max}^{2}}{\sigma_{\min}^{2}},
\end{equation}
which implies that it is always possible to increase the superresolution
precision by aligning the two point sources whenever the Gaussian
point-spread function is not circularly symmetric. On the other hand,
when the Gaussian point-spread function has circular symmetry, the
overlap between the two point-spread functions always remains the
same regardless of the aligning of the two sources, so in this case,
the precision becomes orientation-independent and there is no advantage
in the precision by aligning the two sources.

To illustrate the above results numerically, Fig. \ref{fig:opta}
plots the estimation precision $\hr$ with respect to $\alpha$ and
$\epsilon$ in the limit $r\rightarrow0$. It can be seen that for
a fixed azimuth $\alpha$, the superresolution precision $\hr$ is
maximized at $\epsilon=0$, corresponding to the case of equal photon
intensities. As $\epsilon$ increases, $\hr$ decreases and eventually
vanishes with strong photon intensity imbalance. Moreover, for a given
$\epsilon$, $\hr$ shows clear dependence on the azimuth $\alpha$
between the two point sources, reaching its maximum at
\begin{equation}
\ap=\tan^{-1}\frac{\sigma_{2}^{2}-\sigma_{1}^{2}-\Re}{2\beta\sigma_{1}\sigma_{2}},
\end{equation}
which coincides with the minor axis of the Gaussian point-spread function.

\section{Conclusion}

In this work, we have explored the quantum limits of superresolution
for two incoherent optical point sources in a two-dimensional imaging
system with arbitrary intensity imbalance. By employing the multiparameter
quantum estimation theory, we derived the ultimate precision bounds
for estimating both the distance and the relative orientation (i.e.,
the azimuth) between the two sources. Our results demonstrate that,
unlike classical direct imaging which suffers from Rayleigh’s curse,
quantum strategies can yield nonzero precision even when the distance
between the sources approaches zero. This confirms that the distance
can be resolved beyond the classical diffraction limit through appropriate
quantum measurements.

Moreover, we found that the superresolution precision for distance
estimation has dependence on the relative orientation between two
point sources, which led us to find that the estimation precision
of distance can be significantly enhanced by aligning the two sources
along specific directions and derive the optimal relative azimuth
and the highest estimation precision of the distance between the two
sources. This  opens a new avenue for optimizing superresolution strategies
in practical systems without circular symmetry.

In summary, our findings refine the current understanding of quantum
superresolution and its ultimate capabilities in higher-dimensional
imaging systems, offering both theoretical insight and practical guidance
for the application of quantum superresolution in realistic scenarios.

\section*{ACKNOWLEDGMENTS}

This work is supported by the National Natural Science Foundation
of China (Grant No. 12075323), the Natural Science Foundation of Guangdong
Province of China (Grant No. 2025A1515011440) and the Innovation Program
for Quantum Science and Technology (Grant no. 2021ZD0300702).

\bibliographystyle{apsrev4-2}
\bibliography{superresolution2}

\appendix
\onecolumngrid

\section{Resolution limit of direct imaging}\label{sec:Direct-image}

Suppose the probability distribution of a single photon arriving on
the two-dimensional image plane is given by 
\begin{equation}
p_{{\bf g}}(x,y)=\frac{1-\epsilon}{2}\Lambda(x-X_{1},y-Y_{1})+\frac{1+\epsilon}{2}\Lambda(x-X_{2},y-Y_{2}),\label{eq:dp-1}
\end{equation}
where $(\xk[1,2],\yk[1,2])=\left(\xb\mp\frac{r}{2}{\rm cos}\alpha,\yb\mp\frac{r}{2}{\rm sin}\alpha\right)$.
The photon counting at a small area $dx\times dy$ around the position
$(x,y)$ of the detector is approximately $(N_{1}+N_{2})p_{{\bf g}}(x,y){\rm d}x{\rm d}y$,
and the intensity $\Lambda(x,y)$ is determined by the point-spread
function of imaging system 
\begin{equation}
\Lambda(x,y)=|\psi(x,y)|^{2}.
\end{equation}

Assuming the probability distribution is parametrized by ${\bf \hat{g}}=(\hat{g}_{1},\cdots,\hat{g}_{n})^{{\rm \intercal}}$,
the covariance matrix of any unbiased estimator ${\bf \hat{g}}=(\hat{g}_{1},\cdots,\hat{g}_{n})^{\intercal}$
is lower bounded by the Cramér-Rao inequality \citep{Helstrom1969,Yang2019a,Liu2020},
\begin{equation}
{\rm Cov}[{\bf \hat{g}}]\geq\nt\mathcal{J}^{-1}[{\bf g}],\label{eq:cramer-rao-1}
\end{equation}
where $\nt$ is total number of photons from the two optical sources,
\begin{equation}
\nt=N_{1}+N_{2}.\label{eq:ntot-1}
\end{equation}
The ``$\geq$'' sign represents semi-definite positivity of matrix,
and $\mathcal{J}[{\bf g}]$ is the Fisher information matrix per photon
detection, 
\begin{equation}
\mathcal{J}_{ij}=\int_{-\infty}^{\infty}dy\int_{-\infty}^{\infty}dx\frac{1}{p_{{\bf g}}(x,y)}\frac{\partial p_{{\bf g}}(x,y)}{\partial g_{i}}\frac{\partial p_{{\bf g}}(x,y)}{\partial g_{j}}.\label{eq:classical=00003D000020fisher-1}
\end{equation}
When the total photon number $\nt$ is sufficiently large, the Cramér-Rao
bound inequality can always be asymptotically achieved via the maximum
likelihood estimation strategy (MLE) \citep{Helstrom1969}, i.e.,
$\lim_{\nt\rightarrow\infty}{\rm Cov}[{\bf \hat{g}}]=\nt\mathcal{J}^{-1}[{\bf g}].$

For the task of resolving two point sources, the positions of the
two sources, $(X_{1},Y_{1})$ and $(X_{2},Y_{2})$, or equivalently
the centroid $(\bar{X},\yb)=(\frac{X_{2}+X_{1}}{2},\frac{Y_{2}+Y_{1}}{2})$,
the distance $r$ and the azimuth $\alpha$ are unknown, so the unknown
parameter vector ${\bf g}$ to estimate is ${\bf g}=(\xb,\yb,r,\alpha)^{\top}$.

For the direct imaging with an ideal continuum photodetector on the
image plane, when the distance $r$ is small, one can expand the Fisher
information matrix $\mathcal{J}$ to the second order of $r$, and
the diagonal elements are 
\begin{align}
\mathcal{J}_{11}= & c_{1,0}-\frac{1}{8}r^{2}\bigg\{ a_{2,2}\sin^{2}\alpha+\left(2\epsilon^{2}+1\right)b_{2,0}\sin(2\alpha)+\left(2\epsilon^{2}+1\right)c_{2,0}\cos^{2}\alpha\nonumber \\
 & +2\epsilon^{2}\left[\sin^{2}\alpha\left(c_{1,1}-t_{1,1}\right)-\cos^{2}\alpha f_{1,0}-q_{1,0}\sin(2\alpha)\right]\bigg\},\nonumber \\
\mathcal{J}_{22}= & c_{0,1}-\frac{1}{8}r^{2}\bigg[a_{2,2}\cos^{2}\alpha+\left(2\epsilon^{2}+1\right)\sin\alpha\left(2b_{0,2}\cos\alpha+c_{0,2}\sin\alpha\right)\nonumber \\
 & +2\epsilon^{2}\left(\cos^{2}\alpha\left(c_{1,1}-t_{1,1}\right)-q_{0,1}\sin(2\alpha)-f_{0,1}\sin^{2}\alpha\right)\bigg],\\
\mathcal{J}_{33}= & \frac{1}{4}\epsilon^{2}\left(a_{1,1}\sin(2\alpha)+c_{0,1}\sin^{2}\alpha+c_{1,0}\cos^{2}\alpha\right)+\frac{r^{2}}{64}\bigg[\nonumber \\
 & \left(2-5\epsilon^{2}\right)a_{2,2}\sin^{2}(2\alpha)+8\epsilon^{4}\sin(2\alpha)\left(q_{1,0}\cos^{2}\alpha+q_{0,1}\sin^{2}\alpha\right)\nonumber \\
 & +4\left(2-5\epsilon^{2}\right)\sin(2\alpha)\left(b_{0,2}\sin^{2}\alpha+b_{2,0}\cos^{2}\alpha\right)\nonumber \\
 & +2\left(2-5\epsilon^{2}\right)\left(c_{1,1}\sin^{2}(2\alpha)+c_{2,0}\cos^{4}\alpha+c_{0,2}\sin^{4}\alpha\right)\nonumber \\
 & +4\epsilon^{4}\left(f_{0,1}\sin^{4}\alpha+f_{1,0}\cos^{4}\alpha\right)+6\epsilon^{4}t_{1,1}\sin^{2}(2\alpha)\bigg],\nonumber \\
\mathcal{J}_{44}= & \frac{1}{4}r^{2}\epsilon^{2}\left(c_{1,0}\sin^{2}\alpha+c_{0,1}\cos^{2}\alpha-a_{1,1}\sin(2\alpha)\right),\nonumber 
\end{align}
and the off-diagonal elements of $\mathcal{J}$ are 
\begin{align}
\mathcal{J}_{12}=\mathcal{J}_{21}= & a_{1,1}-\frac{1}{8}r^{2}\bigg[\left(2\epsilon^{2}+1\right)b_{2,0}\cos^{2}\alpha+\left(2\epsilon^{2}+1\right)b_{0,2}\sin^{2}\alpha-2\epsilon^{2}\left(q_{0,1}\sin^{2}\alpha+q_{1,0}\cos^{2}\alpha\right)\nonumber \\
 & +\epsilon^{2}a_{2,2}\sin(2\alpha)+\left(\epsilon^{2}+1\right)c_{1,1}\sin(2\alpha)-2\epsilon^{2}t_{1,1}\sin(2\alpha)\bigg],\nonumber \\
\mathcal{J}_{13}=\mathcal{J}_{31}= & \frac{\epsilon}{2}\left(c_{1,0}\cos\alpha+a_{1,1}\sin\alpha\right)-\frac{\epsilon r^{2}}{16}\bigg[\left(2\epsilon^{2}+1\right)a_{2,2}\sin^{2}\alpha\cos\alpha\nonumber \\
 & +\left(2\epsilon^{2}+1\right)b_{0,2}\sin^{3}\alpha+3\left(2\epsilon^{2}+1\right)b_{2,0}\sin\alpha\cos^{2}\alpha\\
 & +2\left(2\epsilon^{2}+1\right)c_{1,1}\sin^{2}\alpha\cos\alpha+\left(2\epsilon^{2}+1\right)c_{2,0}\cos^{3}\alpha\nonumber \\
 & -2\epsilon^{2}\left(f_{1,0}\cos^{3}\alpha+3\sin\alpha\cos^{2}\alpha q_{1,0}+q_{0,1}\sin^{3}\alpha+3t_{1,1}\sin^{2}\alpha\cos\alpha\right)\bigg],\nonumber \\
\mathcal{J}_{14}=\mathcal{J}_{41}= & \frac{1}{2}r\epsilon\left(a_{1,1}\cos\alpha-c_{1,0}\sin\alpha\right),\nonumber \\
\mathcal{J}_{23}=\mathcal{J}_{32}= & \frac{1}{2}\epsilon\left(a_{1,1}\cos\alpha+c_{0,1}\sin\alpha\right)-\frac{r^{2}\epsilon}{16}\bigg[\left(2\epsilon^{2}+1\right)a_{2,2}\sin\alpha\cos^{2}\alpha+\nonumber \\
 & \left(2\epsilon^{2}+1\right)b_{2,0}\cos^{3}\alpha+3\left(2\epsilon^{2}+1\right)b_{0,2}\sin^{2}\alpha\cos\alpha\nonumber \\
 & +\left(2\epsilon^{2}+1\right)c_{0,2}\sin^{3}\alpha+2\left(2\epsilon^{2}+1\right)c_{1,1}\sin\alpha\cos^{2}\alpha\nonumber \\
 & -2\epsilon^{2}f_{0,1}\sin^{3}\alpha-2\epsilon^{2}q_{1,0}\cos^{3}\alpha-6\epsilon^{2}q_{0,1}\sin^{2}\alpha\cos\alpha\nonumber \\
 & -6\epsilon^{2}t_{1,1}\sin\alpha\cos^{2}\alpha\bigg],\nonumber \\
\mathcal{J}_{24}=\mathcal{J}_{42}= & \frac{1}{2}r\epsilon\left(c_{0,1}\cos\alpha-a_{1,1}\sin\alpha\right),\nonumber \\
\mathcal{J}_{34}=\mathcal{J}_{43}= & \frac{1}{8}r\epsilon^{2}\left[2a_{1,1}\cos(2\alpha)+\left(c_{0,1}-c_{1,0}\right)\sin(2\alpha)\right],\nonumber 
\end{align}
where 
\begin{align}
a_{i,j}= & \begin{aligned}\int_{-\infty}^{\infty}dy\int_{-\infty}^{\infty}dx\frac{\Lambda^{(i,0)}(x,y)\Lambda^{(0,j)}(x,y)}{\Lambda(x,y)}\end{aligned}
,\:i,j=1,2,\nonumber \\
b_{i,j}= & \begin{aligned}\int_{-\infty}^{\infty}dy\int_{-\infty}^{\infty}dx\frac{\Lambda^{(i,j)}(x,y)\Lambda^{(1,1)}(x,y)}{\Lambda(x,y)}\end{aligned}
,\:i,j=0,1,2,\\
c_{i,j}= & \begin{aligned}\int_{-\infty}^{\infty}dy\int_{-\infty}^{\infty}dx\frac{\Lambda^{(i,j)}(x,y)^{2}}{\Lambda(x,y)}\end{aligned}
,\:i,j=0,1,2,\nonumber \\
f_{i,j}= & \begin{aligned}\int_{-\infty}^{\infty}dy\int_{-\infty}^{\infty}dx\frac{\Lambda^{(i,j)}(x,y)^{4}}{\Lambda(x,y)^{3}}\end{aligned}
,\:i,j=0,1,\nonumber \\
q_{i,j}= & \begin{aligned}\int_{-\infty}^{\infty}dy\int_{-\infty}^{\infty}dx\frac{\Lambda^{(0,1)}(x,y)\Lambda^{(1,0)}(x,y)\Lambda^{(i,j)}(x,y)^{2}}{\Lambda(x,y)^{3}}\end{aligned}
,\:i,j=0,1,\nonumber \\
t_{1,1}= & \begin{aligned}\int_{-\infty}^{\infty}dy\int_{-\infty}^{\infty}dx\frac{\Lambda^{(0,1)}(x,y)^{2}\Lambda^{(1,0)}(x,y)^{2}}{\Lambda(x,y)^{3}}\end{aligned}
.\nonumber 
\end{align}

Considering the variance of the distance $r$ may diverge when $r\rightarrow0$,
we use the reciprocal of the variance to quantify the estimation precision
of $r$, the result can be worked out as 
\begin{equation}
\begin{aligned}\mathcal{H}_{r}^{({\rm direct})}=\nt/(\mathcal{J}^{-1})_{33}=\frac{r^{2}\left(1-\epsilon^{2}\right)^{2}\nt}{16}\mathcal{A},\end{aligned}
\end{equation}
where 
\begin{equation}
\mathcal{A}=2a_{2,2}\sin^{2}\alpha\cos^{2}\alpha+2\left(b_{2,0}\cos^{2}\alpha+b_{0,2}\sin^{2}\alpha\right)\sin2\alpha+c_{0,2}\sin^{4}\alpha+c_{1,1}\sin^{2}2\alpha+c_{2,0}\cos^{4}\alpha.
\end{equation}
As the precision $\mathcal{H}_{r}^{({\rm direct})}$ drops quadratically
as the distance $r$ decreases, it vanishes when $r$ approaches zero,
implying no information about $r$ can be obtained when the two point
sources are too close to each other. This phenomenon is called Rayleigh’s
curse \citep{Tsang2016}.

\section{Precision of quantum superresolution}

\subsection{Multiparameter quantum estimation theory for superresolution}\label{subsec:Derivation=00003D000020f}

The quantum parameter estimation theory provides a mathematical tool
to address Rayleigh's curse for two closely spaced optical point sources.
In general, it takes into account the parameters in a quantum system
and further optimizes the estimation precision over all possible measurement
bases in addition to the optimization of estimation strategies involved
in the classical Cramér-Rao bound. The precision of quantum estimation
is characterized by the quantum Cramér-Rao bound, 
\begin{equation}
{\rm Cov}[{\bf \hat{g}}]\geq\nt[-1]\mathcal{Q}_{ij}^{-1}[{\bf g}],
\end{equation}
where $\mathcal{Q}$ is the quantum Fisher information matrix, defined
as 
\begin{equation}
\mathcal{Q}_{ij}(\hat{\rho})=\frac{1}{2}{\rm Tr}[\hro\{\hl{g_{i}},\hl{g_{j}}\}],\,,\forall g_{i},g_{j}\in{\bf g},\label{eq:fisher0-1}
\end{equation}
where $\{\cdot,\cdot\}$ denotes the anticommutator and $\hl{g_{i}}$
is the symmetric logarithmic derivative (SLD) of the density operator
$\hro$ with respect to the parameter $g_{i}$, 
\begin{equation}
\frac{1}{2}(\hl{g_{i}}\hro+\hro\hl{g_{i}})=\partial_{g_{i}}\hro.\label{eq:sld-1}
\end{equation}
By decomposing the density matrix $\hro$ in its orthogonal eigenvectors
as 
\begin{equation}
\hro=\sum_{k}\lambda_{k}|\lambda_{k}\rangle\langle\lambda_{k}|,
\end{equation}
a SLD $\hl{g_{i}}$ can be expressed as 
\begin{equation}
\hl{g_{i}}=\sum_{\lambda_{k}+\lambda_{h}\neq0}\frac{2\langle\lambda_{k}|\partial_{g_{i}}\hat{\rho}|\lambda_{h}\rangle}{\lambda_{k}+\lambda_{h}}\ket{\lambda_{k}}\bra{\lambda_{h}}.
\end{equation}
Since the density matrix \eqref{eq:rho} of two point sources has
rank 2, the quantum Fisher information matrix is more conveniently
calculated within the support of $\hro$, 
\begin{equation}
\mathcal{Q}_{ij}(\hat{\rho})=\sum_{\lambda_{k}\neq0}\frac{4\langle\lambda_{k}|\partial_{g_{i}}\hat{\rho}\partial_{g_{j}}\hat{\rho}|\lambda_{k}\rangle}{\lambda_{k}}+\sum_{\lambda_{k},\lambda_{h}\neq0}2\left(\frac{1}{\lambda_{k}+\lambda_{h}}-\frac{1}{\lambda_{k}}-\frac{1}{\lambda_{h}}\right)\langle\lambda_{h}|\partial_{g_{i}}\hat{\rho}|\lambda_{k}\rangle\langle\lambda_{k}|\partial_{g_{j}}\hat{\rho}|\lambda_{h}\rangle.\label{eq:fisher=00003D000020exp}
\end{equation}

In fact, the quantum Cramér-Rao bound for multiparameter estimation
is not always achievable due to the potential non-commutativity between
the optimal measurements for different parameters. But if one is interested
in the overall estimation precision of different parameters, a positive-semidefinite
weight matrix $W$ can be imposed on the unknown parameters generally
and the matrix form of quantum Cramér-Rao's bound can be simplified
to a scalar form, 
\begin{equation}
{\rm Tr}(WC)\geq\nt[-1]{\rm Tr}(W\mathcal{Q}^{-1}).\label{eq:scalar-1}
\end{equation}
By introducing the weight matrix $W$, we can assign different weights
to different unknown parameters. Furthermore, the quantum Cramér-Rao
bound for the above scalar form can be saturated if 
\begin{equation}
{\rm Tr}(\hro[\hl{g_{i}},\hl{g_{j}}])={\rm ImTr}(\hro\hl{g_{i}}\hl{g_{j}})=0,\label{eq:compatibility-1}
\end{equation}
where $[\cdot,\cdot]$ denotes the commutator, by recognizing that
the quantum Cramér-Rao bound is equivalent to the Holevo bound when
this condition is satisfied and the latter is always achievable asymptotically
with a sufficiently large number of states \citep{Ragy2016,Suzuki2016}.
Note that for our current quantum superresolution problem, the compatibility
condition \eqref{eq:compatibility-1} is always fulfilled due to the
reality of point-spread function which makes the symmetric logarithmic
derivatives $\hl{g_{i}}$ and $\hl{g_{j}}$ also real according to
Eq. \eqref{eq:sld-1}. So the quantum Cramér-Rao bound provides an
achievable precision limit for the estimation of the distance between
two incoherent point sources allowed by quantum mechanics \citep{Ragy2016,Rehacek2017}.

\subsection{Quantum estimation precision of source distance}\label{subsec:2dqf}

Building upon the pioneering quantum superresolution framework for
one-dimensional imaging systems developed by Tsang et al. \citep{Tsang2016},
we extend the superresolution theory to two incoherent optical point
sources with arbitrary intensities in two-dimensional imaging systems.
The quantum state of a single detected photon can be described by
the density operator 
\begin{equation}
\begin{aligned}\hro= & \frac{1-\epsilon}{2}\nei[x]{\xk[1]}\nei[y]{\yk[1]}\ksi\bsi\ei[x]{\xk[1]}\ei[y]{\yk[1]}\\
 & +\frac{1+\epsilon}{2}\nei[x]{\xk[2]}\nei[y]{\yk[2]}\ksi\bsi\ei[x]{\xk[2]}\ei[y]{\yk[2]},
\end{aligned}
\label{eq:ro}
\end{equation}
which depends on the parameters ${\bf g}=(\xb,\yb,r,\alpha)^{\top}$.

The support of the density matrix $\hro$ \eqref{eq:ro} is spanned
by two eigenstates associated with nonzero eigenvalues, 
\begin{align}
|\lambda_{1,2}\rangle & =\frac{1}{\sqrt{\mathscr{Q}_{1,2}}}\times\{\nei[y]{Y_{1}}\nei[x]{X_{1}}\ksi\mp\frac{\left(\sqrt{\epsilon^{2}+\delta^{2}\left(1-\epsilon^{2}\right)}\mp\epsilon\right)}{\delta(1-\epsilon)}\nei[x]{\xk[2]}\nei[y]{\yk[2]}\ksi\},\label{eq:es}
\end{align}
where $\mathscr{Q}_{1,2}$ are the normalization constants, 
\begin{align}
\mathscr{Q}_{1,2} & =\left(1-\delta^{2}\right)\left(1+\frac{\delta^{2}\left(1-\delta^{2}\right)(1-\epsilon)^{2}}{\left(\sqrt{\delta^{2}+\epsilon^{2}-\delta^{2}\epsilon^{2}}\pm\delta^{2}(1-\epsilon)\pm\epsilon\right)^{2}}\right),
\end{align}
and $\delta$ is the overlap between two point-spread functions 
\begin{equation}
\delta\equiv\left\langle \psi_{1}\mid\psi_{2}\right\rangle =\int_{-\infty}^{\infty}dx\int_{-\infty}^{\infty}dy\psi_{1}^{\ast}(x,y)\psi_{2}(x,y)\neq0.
\end{equation}
The eigenvalues are 
\begin{equation}
\lambda_{1,2}=\frac{1}{2}\left(1\mp\sqrt{\epsilon^{2}+\delta^{2}\left(1-\epsilon^{2}\right)}\right).\label{eq:ev}
\end{equation}
Since the two point-spread functions are real, the overlap can be
simplified to 
\begin{equation}
\delta=\av[{\nei[x]{r{\rm cos}\alpha}\nei[y]{r{\rm sin}\alpha}}]=\av[{\cos(r\hp[r])}].
\end{equation}

According to Eq. \eqref{eq:fisher=00003D000020exp}, the quantum Fisher
information matrix with respect to unknown parameters ${\bf g}=\left(\xb,\yb,r,\alpha\right)^{\intercal}$
can be derived by plugging the eigenvalues \eqref{eq:ev} and eigenstates
\eqref{eq:es} of the density matrix $\hro$ into Eq. \eqref{eq:fisher=00003D000020exp},
and the elements of the quantum Fisher information matrix $\mathcal{Q}$
turn out to be 
\begin{align}
\mathcal{Q}_{11}= & 4\kappa_{x}-4\left(1-\epsilon^{2}\right)\gamma_{x}^{2},\nonumber \\
\mathcal{Q}_{12}=\mathcal{Q}_{21}= & 4\eta-4\left(1-\epsilon^{2}\right)\gamma_{x}\gamma_{y},\nonumber \\
\mathcal{Q}_{13}=\mathcal{Q}_{31}= & 2\epsilon\left(\eta\sin\alpha+\cos\alpha\kappa_{x}\right),\nonumber \\
\mathcal{Q}_{14}=\mathcal{Q}_{41}= & 2r\epsilon\left(\eta\cos\alpha-\sin\alpha\kappa_{x}\right),\nonumber \\
\mathcal{Q}_{22}= & 4\kappa_{y}-4\left(1-\epsilon^{2}\right)\gamma_{y}^{2},\nonumber \\
\mathcal{Q}_{23}=\mathcal{Q}_{32}= & 2\epsilon\left(\eta\cos\alpha+\sin\alpha\kappa_{y}\right),\label{eq:qfim2d}\\
\mathcal{Q}_{24}=\mathcal{Q}_{42}= & 2r\epsilon\left(\cos\alpha\kappa_{y}-\eta\sin\alpha\right),\nonumber \\
\mathcal{Q}_{33}= & \kappa_{r},\nonumber \\
\mathcal{Q}_{34}=\mathcal{Q}_{43}= & \frac{r}{2}\left(2\eta\cos(2\alpha)-\sin(2\alpha)\kappa_{x}+\sin(2\alpha)\kappa_{y}\right),\nonumber \\
\mathcal{Q}_{44}= & r^{2}\kappa_{r^{\perp}},\nonumber 
\end{align}
where 
\begin{align}
\kappa_{x}= & \avg[x][][][2],\kappa_{y}=\avg[y][][][2],\eta=\av[{\hp[x]\hp[y]}],\nonumber \\
\kappa_{r^{\perp}}= & \avg[r^{\perp}][][][2]=\kappa_{x}\sin^{2}\alpha+\kappa_{y}\cos^{2}\alpha-\eta\sin(2\alpha),\\
\kappa_{r}= & \avg[r][][][2]=\eta\sin(2\alpha)+\cos^{2}\alpha\kappa_{x}+\sin^{2}\alpha\kappa_{y},\nonumber \\
\gamma_{x}= & i\av[{\hp[x]\nei[x]{r{\rm cos}\alpha}\nei[y]{r{\rm sin}\alpha}}]=\av[{\hp[x]\sin(r\hp[r])}],\nonumber \\
\gamma_{y}= & i\av[{\hp[y]\nei[x]{r{\rm cos}\alpha}\nei[y]{r{\rm sin}\alpha}}]=\av[{\hp[y]\sin(r\hp[r])}].\nonumber 
\end{align}

In the present quantum superresolution problem, the primary parameter
of interest is the distance $r$. To focus on the estimation precision
of $r$, we introduce a weight matrix 
\begin{equation}
W=\begin{bmatrix}0 & 0 & 0 & 0\\
0 & 0 & 0 & 0\\
0 & 0 & 1 & 0\\
0 & 0 & 0 & 0
\end{bmatrix},
\end{equation}
so that ${\rm Tr}(WC)$ is the variance of $r$ which is bounded by
$\nt[-1]{\rm Tr}(W\mathcal{Q}^{-1})$, where $C$ is the covariance
matrix of the parameters ${\bf g}=(\xb,\yb,r,\alpha)^{\top}$. By
inverting the quantum Fisher information matrix $Q$, the estimation
precision of the distance $r$ can be determined as 
\begin{equation}
\hr=\nt/(\mathcal{Q}^{-1})_{33}=\frac{\hrn}{\hrd},\label{eq:hr}
\end{equation}
where the numerator is 
\begin{equation}
\begin{aligned}\hrn & =\nt\left(1-\epsilon^{2}\right)\left(\kappa_{x}\kappa_{y}-\eta^{2}\right)\times\\
 & \left[\kappa_{x}\kappa_{y}-\eta^{2}+2\eta\gamma_{x}\gamma_{y}-\kappa_{x}\gamma_{y}^{2}-\gamma_{x}^{2}\kappa_{y}\right],
\end{aligned}
\end{equation}
and the denominator is 
\begin{equation}
\begin{aligned}\hrd= & \left(\kappa_{x}\kappa_{y}-\eta^{2}\right)\kappa_{r^{\perp}}\\
 & +\gamma_{x}^{2}\left[\epsilon^{2}\left(\eta\sin\alpha-\kappa_{y}\cos\alpha\right){}^{2}-\kappa_{r^{\perp}}\kappa_{y}\right]\\
 & +\gamma_{x}\gamma_{y}\left[\epsilon^{2}\left(\kappa_{x}\kappa_{y}-\eta^{2}\right)\sin(2\alpha)+2\eta\kappa_{r^{\perp}}\left(1-\epsilon^{2}\right)\right]\\
 & +\gamma_{y}^{2}\left[\epsilon^{2}\left(\eta\cos\alpha-\kappa_{x}\sin\alpha\right){}^{2}-\kappa_{r^{\perp}}\kappa_{x}\right].
\end{aligned}
\end{equation}

We focus on the estimation precision $\hr$ in the sub-Rayleigh regime,
where the distance $r$ is much smaller than the classical diffraction
limit. To investigate the quantum limit of estimating the distance
$r$ in this regime, we consider the precision $\hr$ in the asymptotic
case $r\rightarrow0$, which is reduced to 
\begin{equation}
\dlm{\hr}=\frac{\text{\ensuremath{\nt}}\left(1-\epsilon^{2}\right)\left(\kappa_{x}\kappa_{y}-\eta^{2}\right)}{\kappa_{r^{\perp}}}.\label{eq:hds-1}
\end{equation}
This result demonstrates that, while direct imaging fails with vanishing
precision in the sub-Rayleigh regime, the quantum strategy achieves
non-zero precision, thus beating Rayleigh’s curse.

For two Gaussian point-spread functions, the quantities in Eq. \eqref{eq:hds-1}
become 
\begin{equation}
\begin{aligned}\kappa_{x}=\frac{1}{4\sigma_{1}^{2}\left(1-\beta^{2}\right)},\\
\kappa_{y}=\frac{1}{4\sigma_{2}^{2}\left(1-\beta^{2}\right)},\\
\eta=-\frac{\beta}{4\sigma_{1}\sigma_{2}\left(1-\beta^{2}\right)}.
\end{aligned}
\end{equation}
The estimation precision for the distance $r$ in the limit $r\rightarrow0$,
can be simplified to 
\begin{equation}
\dlm{\hr}=\frac{\left(1-\epsilon^{2}\right)\nt}{4\left(\beta\sigma_{2}\sigma_{1}\sin(2\alpha)+\sigma_{2}^{2}\sin^{2}(\alpha)+\sigma_{1}^{2}\cos^{2}(\alpha)\right)},\label{eq:hdg-1}
\end{equation}
which shows $\hr$ keeps nonzero when $r$ is zero.

\subsection{Coordinate invariance of $\protect\hr$}\label{subsec:The-Coordinate-Invariance}

While the precision limit $\hr$ \eqref{eq:hds-1} seems dependent
on the azimuth $\alpha$ which can change with the coordinate system,
$\hr$ should be invariant with any change of the coordinate system
as it is an inherent property of the two point sources irrelevant
to what coordinate system we choose.

To verify the coordinate-independence of the $\hr$, we first note
that the factor $\kappa_{x}\kappa_{y}-\eta^{2}$ in the numerator
of $\hr$ \eqref{eq:hds-1} is actually the determinant of the following
matrix, 
\begin{equation}
\Pi=\left[\begin{array}{cc}
\kappa_{x} & \eta\\
\eta & \kappa_{y}
\end{array}\right],
\end{equation}
with $\kappa_{x}=\avg[x][][][2],\kappa_{y}=\avg[y][][][2],\eta=\av[{\hp[x]\hp[y]}]$.
Under a global rotation of the coordinate system by an angle $\theta$,
the momentum operators $\hp[x],\hp[y]$ are transformed as 
\begin{equation}
\left[\begin{array}{c}
\hp[x][\prime]\\
\hp[y][\prime]
\end{array}\right]=R(\theta)\left[\begin{array}{c}
\hp[x]\\
\hp[y]
\end{array}\right],
\end{equation}
where 
\begin{equation}
R(\theta)=\left[\begin{array}{c}
{\rm cos}\theta\\
{\rm sin}\theta
\end{array}\begin{array}{c}
{\rm -sin}\theta\\
{\rm cos}\theta
\end{array}\right].
\end{equation}
So the matrix $\Pi$ becomes 
\begin{equation}
\Pi^{\prime}=\left[\begin{array}{cc}
\kappa_{x}^{\prime} & \eta^{\prime}\\
\eta^{\prime} & \kappa_{y}^{\prime}
\end{array}\right]=R(\theta)^{\intercal}\Pi R(\theta),
\end{equation}
in the rotated basis.

Since a similarity transformation preserves the determinant of a matrix,
we have 
\begin{equation}
{\rm det(\Pi^{\prime})=det(\Pi)},
\end{equation}
so the quantity $\kappa_{x}\kappa_{y}-\eta^{2}$ remains invariant
with the rotation of the coordinate system.

For the denominator of the precision $\hr$ in Eq. \eqref{eq:hds-1},
we show that this is equal to the mean square of the momentum along
the direction orthogonal to the displacement between the two point
sources, i.e., 
\begin{equation}
\kappa_{r^{\perp}}=\avg[r^{\perp}][][][2]=\kappa_{x}\sin^{2}\alpha+\kappa_{y}\cos^{2}\alpha-\eta\sin2\alpha,
\end{equation}
where 
\begin{equation}
\hp[r^{\perp}]=\hp[y]\cos\alpha-\hp[x]\sin\alpha.
\end{equation}

As the displacement between the two point sources is physically inherent
to the sources, irrelevant to the choice of coordinate system, the
momentum operator $\hp[r^{\perp}]$ in the direction orthogonal to
the displacement between the two sources is a physical observable
independent of the coordinate system, therefore, the mean square of
$\hp[r^{\perp}]$ is physically determined and remains invariant under
any coordinate rotation.

So, we can see that both the numerator and denominator of the precision
$\hr$ \eqref{eq:hds-1} are coordinate-invariant quantities: the
numerator is the determinant of the matrix $\Pi$ composed of the
mean of the products of two momentum operators, and the denominator
is the mean square of the momentum along the direction orthogonal
to the displacement between the two sources. Consequently, $\hr$
is independent of the choice of the coordinate system and depends
solely on the intrinsic geometric relation between the point-spread
functions of the two sources. 
\end{document}